\documentclass[runningheads]{cl2emult}

\usepackage{makeidx}  
\usepackage{graphicx} 
\usepackage{subeqnar} 
\usepackage{multicol} 
\usepackage{cropmark} 
\usepackage{eso}      
\makeindex            

\begin{document}
\title*{Synthetic Stellar Clusters for Pop III}
\toctitle{Synthetic Stellar Clusters for Pop III}
%
%
\titlerunning{Synthetic Stellar Clusters for Pop III}
%
\author{G. Raimondo\inst{1}
\and E. Brocato\inst{1}
\and S. Cassisi\inst{1}
\and V. Castellani\inst{2}
}
\authorrunning{G. Raimondo et al.}
%
%
\institute{Osservatorio Astronomico di Teramo, Teramo, Italy 
\and Universit\`a di Pisa, dip. di Fisica, Universit\`a di Pisa, Pisa, Italy}

\maketitle              

\begin{abstract}
We present preliminary results of an incoming theoretical work
concerning the integrated properties of the Population III clusters of
stars.  On the basis of synthetic Color-Magnitude Diagrams, we provide
a grid of optical and near-IR colors of Simple Stellar Populations
with very low metallicity (Z=10$^{-10}$ and Z=10$^{-6}$) and age which
spans from 10 Myr to 15 Gyr.  A comparison with higher metallicities
up to 0.006 is also shown, disclosing sizable differences in the CMD
morphology, integrated colors and Spectral Energy Distribution (SED).
\end{abstract}

\section{Metal deficient simple star populations}

In the last two decades a great theoretical effort has been devoted to
investigate the integrated properties of metal rich and metal poor
star populations in the local universe (Bressan et al. 1994; Leitherer
et al. 1996) providing a modellization of integrated colors and
spectra as a function of the IMF, age, chemical content etc...
However, till now a complete evaluation of the integrated properties
for first generation of stars which should directly form from matter
emerged from the big bang is still missing.

As a part of a project devoted to investigate the integrated
properties of stellar clusters in a homogeneous and complete framework
(Brocato et al. 1999, Brocato et al. in preparation), we approach
here the analysis of the spectro-photometric behavior of metal
deficient star systems.  By relying on the extended grid of
evolutionary models for very metal poor stars (Z=10$^{-10}$ and
Z=10$^{-6}$ with Y=0.23), including evolutionary phases from the MS up
to the double shell burning, previously calculated (Cassisi et
al. 1996; Cassisi and Castellani, 1993) we performed synthetic
population models for a wide range of age from 10 Myr to 15 Gyr
adopting a Salpeter mass distribution (IMF).

Selected integrated colors in the optical and near-IR bands are
reported in Fig. 1$a$. As expected from the synthetic CM Diagrams, the
integrated colors become bluer and bluer as the metallicity decreases
for the older cluster. Moreover, the younger clusters appear to show
similar colors even with very different assumptions on metallicity (up
to Z=10$^{-4}$), since the hotter and bluer stars dominate the
spectral energy distribution.

\begin{figure}[t]
\vspace{-1cm}
\center{
\includegraphics[width=.4\textwidth]{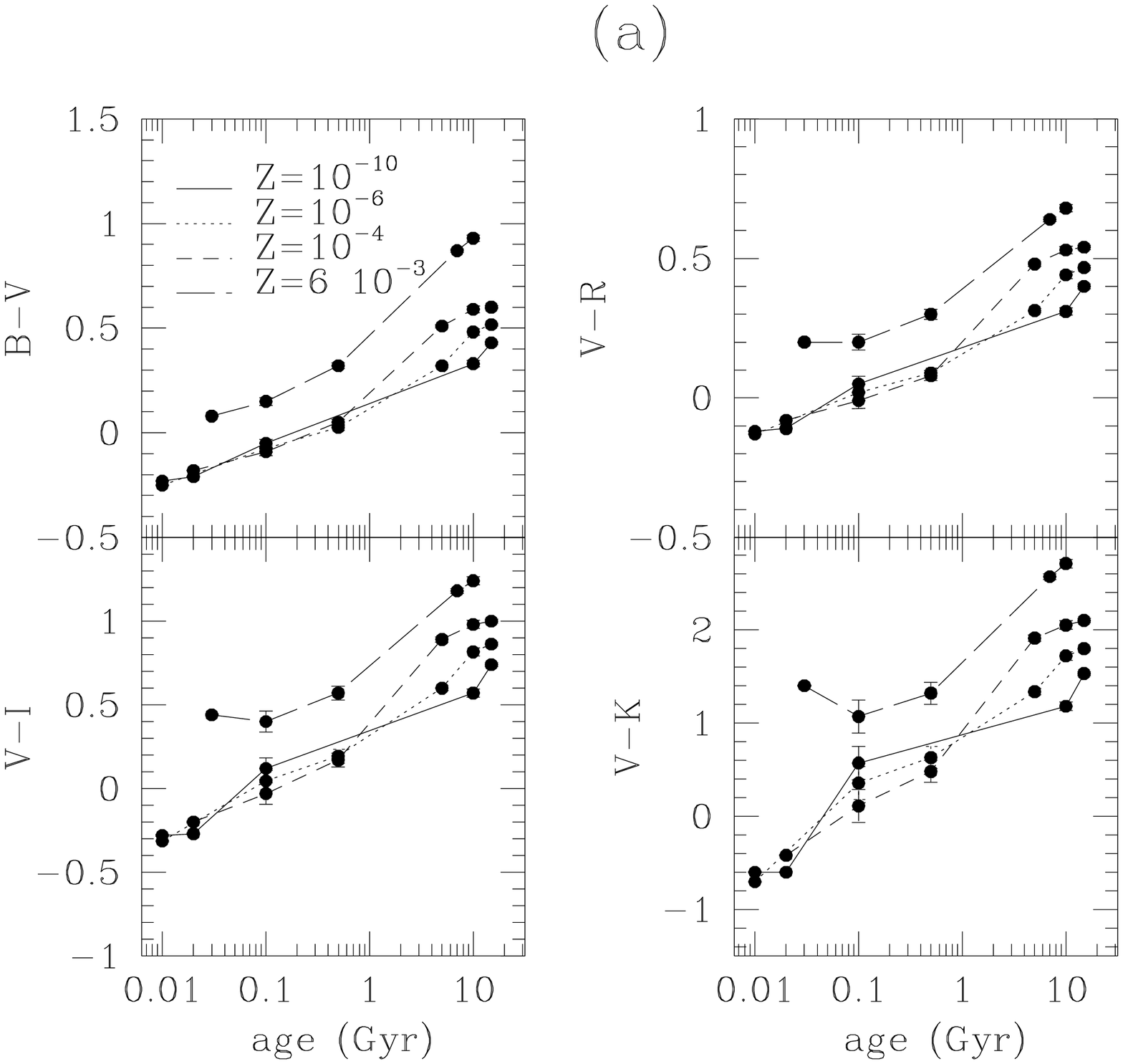}
\hspace{0.9cm}
\includegraphics[width=.5\textwidth]{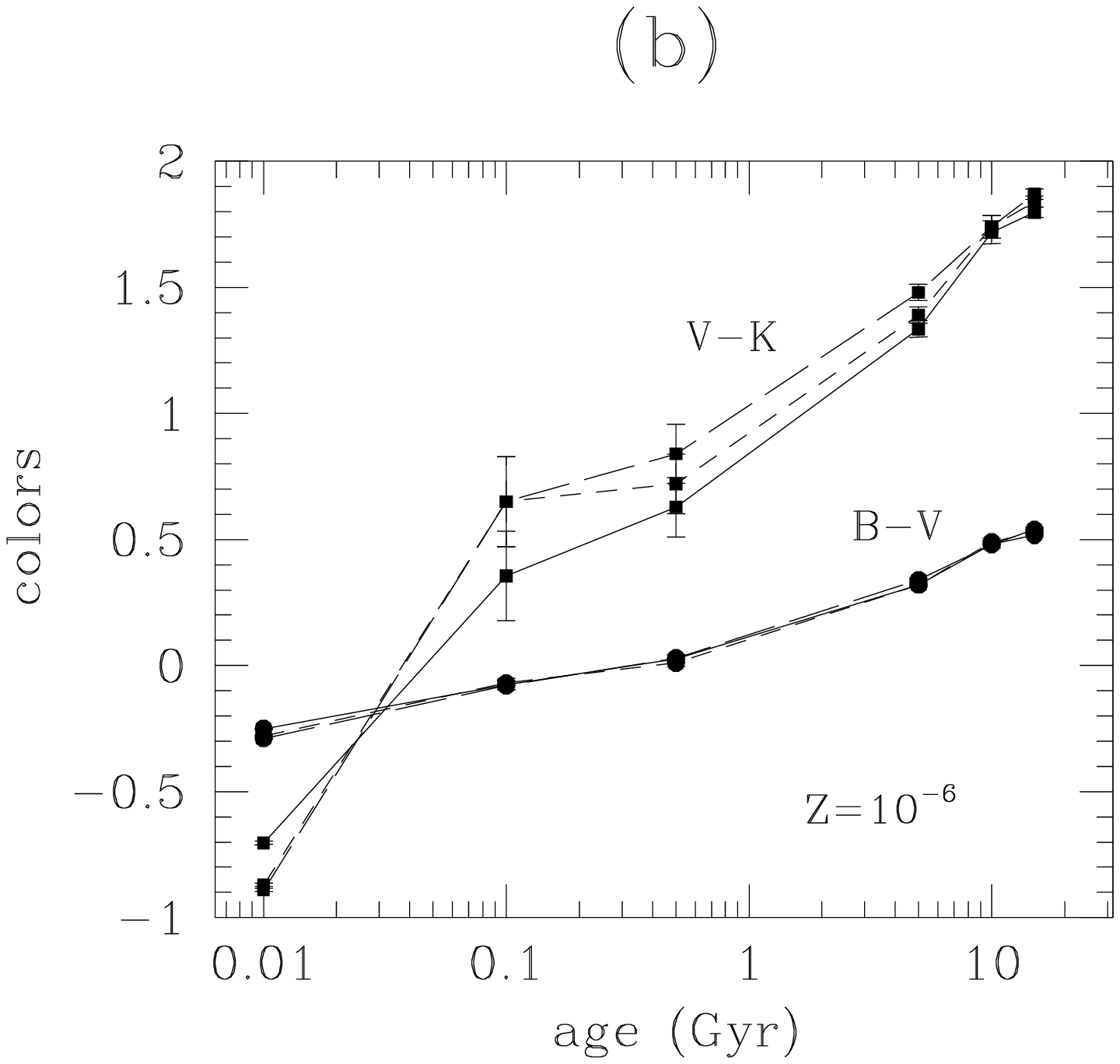}
}
\caption[]{$(a)$ Integrated colors for four different star chemical contents: 
Z=10$^{-10}$ (solid line), Z=10$^{-6}$ (dotted line), Z=10$^{-4}$ (dashed 
line) and Z=6$\cdot$10$^{-3}$ (long dashed line). 
$(b)$ The variability in B$-$V and 
V$-$K colors calculated assuming: $i)$ a power law with a Salpeter slope (solid line), 
$ii)$ the Larson form with M$_{p}\sim$ 1.5M$_{\odot}$ (dashed line) and $iii)$ the Larson form with 
M$_{p}\sim$ 3M$_{\odot}$ (long dashed line). Z=10$^{-6}$ is adopted.}
\label{eps1}
\end{figure}

\section{Effect of the IMF on the integrated colors}

The long standing debate on the universality of the IMF is of
particular relevance in the case of first star generation. Both theory
and observation seem to suggest that the IMF has a typical power low
at more massive stars and the deviation from this form should be
confined at the lower part of the mass distribution.
The thermo-dynamical conditions of primordial gas should imply a
larger value of ``mass-scale'' (M$_{p}$) (Larson 1998) 
compared to the present one (some
tenth of solar mass), causing a deficiency of less massive stars.  We
investigate the effect of changing the IMF on the integrated colors
adopting the analytic form suggested by Larson (1998): a Salpeter
power law at larger masses, which peaks at M$_{p}$ and falls
exponentially at lower masses.  We find (Fig. 1$b$) that the B$-$V
color does not change for the three assumptions on the IMF. The V$-$K
color shows the most evident variations, since it is strongly affected
by the differences of the number of stars in the mass intervals.


\clearpage
\addcontentsline{toc}{section}{Index}
\flushbottom

\end{document}